\documentclass[12pt]{iopart}
\usepackage{graphicx}
\usepackage{iopams}  

\expandafter\let\csname equation*\endcsname\relax
\expandafter\let\csname endequation*\endcsname\relax
\usepackage{amsmath}
\usepackage{cite}
\begin{document}

\title{Revival of Silenced Echo and Quantum Memory for Light}

\author{V. Damon, M. Bonarota, A. Louchet-Chauvet, T. Chaneli\`ere, J.-L. Le Gou\"et}

\address{Laboratoire Aim\'e Cotton, CNRS-UPR 3321, Univ. Paris-Sud, B\^at. 505, 91405 Orsay cedex, France}
\ead{jean-louis.legouet@lac.u-psud.fr}
\begin{abstract}
We propose an original quantum memory protocol. It belongs to the class of rephasing processes and is closely related to two-pulse photon echo. It is known that the strong population inversion produced by the rephasing pulse prevents the plain two-pulse photon echo from serving as a quantum memory scheme. Indeed gain and spontaneous emission generate prohibitive noise. A second $\pi$-pulse can be used to simultaneously reverse the atomic phase and bring the atoms back into the ground state. Then a secondary echo is radiated from a non-inverted medium, avoiding contamination by gain and spontaneous emission noise. However, one must kill the primary echo, in order to preserve all the information for the secondary signal. In the present work, spatial phase mismatching is used to silence the standard two-pulse echo. An experimental demonstration is presented.       
\end{abstract}
%Uncomment for PACS numbers title message
% Keywords required only for MST, PB, PMB, PM, JOA, JOB? 
%\vspace{2pc}
%\noindent{\it Keywords}: Article preparation, IOP journals
% Uncomment for Submitted to journal title message
%\submitto{\JPA}
% Comment out if separate title page not required
\maketitle

\section{Introduction}
The interaction of quantum light with an ensemble of atoms has given rise to intense research efforts during the past decade. The emblematic quantum memory challenge includes the conversion of a quantum state of light into an ensemble state of matter and the retrieval of a restored state of light. Light operates as a probe of the ensemble entangled state. Electromagnetically induced transparency (EIT) combined with the DLCZ generation of narrow-band heralded photons led to convincing experimental demonstrations in atomic vapors~\cite{Chaneliere2005,Eisaman2005,appel2008}. However one cannot help but notice the small bandwidth offered by EIT-based protocols. Indeed the spectrum of the incoming pulse of light has to fit the transparency window that is opened within the homogeneous width of the atomic transition. This practically limits the bandwidth to the megahertz range. To break this barrier one may map the signal spectral components over the inhomogeneous width of the absorption line, according to the paradigm of the well known photon echo scheme. In this context, solid state materials such as rare-earth ion doped crystals (REIC) offer a prime alternative to atomic vapors since they combine the absence of motion with long coherence lifetime and large inhomogeneous width.

Although successful as a classical signal storage technique, photon echo does not work right away as a quantum memory procedure, as pointed out by several authors~\cite{ruggiero2009,sangouard2010,ledingham2010}. For instance, the two-pulse echo (2PE) suffers from the population inversion produced by the rephasing pulse. To efficiently reverse the phase of atomic coherences and get them phased together at a later time, this pulse has to simultaneously promote the atoms to the upper level of the optical transition. Working in a gain regime, detrimental to fidelity, the inverted medium also relaxes by spontaneous emission (SE), which further increases the intrinsic noise and make this scheme improper to the recovery of the initial quantum state of light. Last but not least, due to the spatial phase matching requirement, the 2PE signal propagates along the same direction as the driving fields. Hence, when composed of a few photons, the echo gets buried in the free induction decay (FID) long tail of the rephasing pulse~\cite{ruggiero2009}.

To adapt the photon echo to quantum memory requirements, one has to get rid of massive population inversion. Original protocols such as the controlled reversible inhomogeneous broadening (CRIB)~\cite{Moiseev2001,nilsson2005,kraus2006,sangouard2007}, the gradient echo memory (GEM)~\cite{alexander2006,Hetet2008,Hetet2008bis,Longdell2008,hedges2010,Lauritzen2010} and the atomic frequency comb (AFC)\cite{Riedmatten2008,afzelius2009} avoid population inversion. Either the phase reversal is produced by an external electric or magnetic field (CRIB and GEM), or rephasing just results from the initial preparation of the inhomogeneously broadened distribution (AFC). Those techniques have proved very successful in terms of efficiency \cite{Bonarota2010,hedges2010}, multimode capacity~\cite{usmani2010,bonarota2011}, and quantum fidelity~\cite{saglamyurek2011,clausen2011}. However they all require a rather complex preparation step. In addition, by removing atoms from the absorption band, the preparation step reduces the intrinsic trapping capability of the medium.

In the present paper, returning to 2PE basics, we propose an alternative approach, free from any preparation step. As already noticed by J. J. Longdell~\cite{ledingham2010,longdell2010,mcauslan2011}, the issue is the population inversion, and the resulting gain and SE noise, rather than the intense pulse by itself. A second intense pulse is able to bring the atoms back to the ground state, thus suppressing the undesired noise, and to simultaneously reverse the phase of the atoms, making them emit a secondary echo. The whole question is how to silence the primary 2PE and to preserve the integrity of the captured information until the emission of the secondary echo. Longdell~\cite{longdell2010,mcauslan2011} proposed to kill the primary echo by Stark-effect-induced interference~\cite{meixner1992,chaneliere2008}. We here propose to rely on spatial phase mismatching, which eliminates the need for an external field. This is explained in section~\ref{protocol} where we describe the storage protocol. In section~\ref{Rephasing pulses} we discuss about rephasing the Bloch vectors with a pair of adiabatic rapid passages. Experimental results are presented in section~\ref{experimental}.    
   
\section{Rephasing in the ground state with silenced primary echo}\label{protocol}
\begin{figure}[h!]
\centering{\includegraphics[width=10cm]{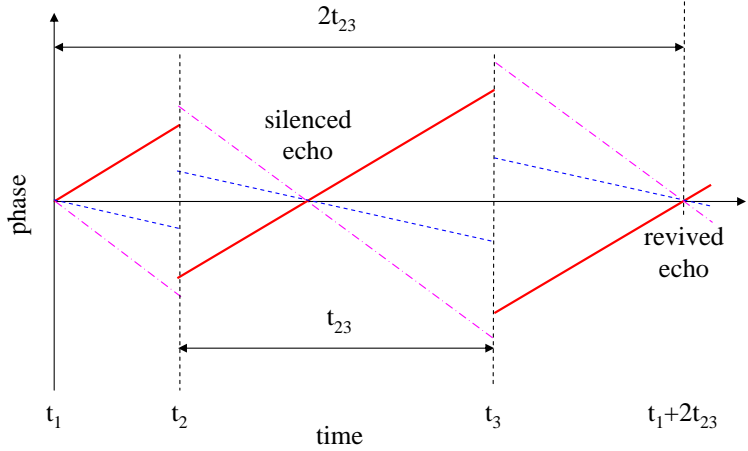}}
\caption{(color online) Revival Of Silenced Echo. Excitation at time $t_1$ gives rise to atomic coherences. Departing from their initial common phase, the atomic coherences evolve at different rates, depending on their detuning from a reference. Rephasing pulses are shone at times $t_2$ and $t_3$. The atomic coherences get phased together at time $t_1+2t_{12}$ but the primary echo is silenced by spatial phase mismatching. The echo is revived at time $t_2+2t_{23}$.}
\label{fig:ROSE}
\end{figure}

Let us consider a simple 2PE scheme, operating on an ensemble of two-level atoms. First a weak pulse, carrying the information to be stored, shines the storage medium at time $t_1$. Then a strong pulse hits the medium at time $t_2$, rotating the Bloch vector by an angle of $\pi$. The $\pi$-rotation simultaneously reverses the inhomogeneous phase shift and promotes the atoms to the upper level. The atomic coherences get phased together again at time $t_e=t_1+2t_{12}$, where $t_{ij}=t_j-t_i$, and radiate an echo signal. The optical thickness must be large enough to allow for efficient capture of the initial signal pulse. In such a medium, population inversion by the rephasing pulse yields important gain and SE. Both phenomena affect the fidelity of the information recovery. To bring the atoms back to the ground state, one just has to apply a second $\pi$-pulse at time $t_3>t_e$. As shown in Fig.~\ref{fig:ROSE}, a secondary echo is then emitted at time $t_{e'}=t_1+2t_{23}$ from ground state atoms, free of gain and SE noise. However, this simple procedure only enables us to recover a part of the stored information. Another part has already been carried away by the primary echo at time $t_e$. To avoid this loss of information one has to silence the primary echo. 

We rely on spatial phase mismatching to kill the primary echo. The pulses, directed along wavevectors $\vec{k}_1$ and $\vec{k}_2$, give rise to echo emission in direction $\vec{k}_e=2\vec{k}_2-\vec{k}_1$, provided $k_e=|\vec{k}_e|$ is close to $k=|\vec{k}_1|=|\vec{k}_2|$. More precisely, the condition reads: $\left(k_e-k\right)L<<\pi$, where $L$ stands for the medium thickness. Indeed, the radiated field must match the spatial phase of the radiating atomic macroscopic polarization. If $\left(k_e-k\right)L>\pi$, no echo is emitted at time $t_e$ but the macroscopic polarization does survive. The violation of the phase matching condition does not affect atomic coherences. The spectral phase shift is reversed anyway at time $t_3$ and, despite the absence of the primary echo, a secondary echo can be emitted at time $t_{e'}=t_1+2t_{23}$ provided $\vec{k}_{e'}=2\vec{k}_3-\vec{k}_e=\vec{k}_1+2\left(\vec{k}_3-\vec{k}_2\right)$ satisfies the phase matching condition. If $\vec{k}_3=\vec{k}_2$, emission takes place in the same $\vec{k}_1$ direction as the initial signal, whatever the common wavevector direction of the two rephasing pulses. This occurs for instance when the rephasing pulses counterpropagate with the incoming signal, a configuration that strictly forbids the 2PE emission. We coin this process Revival Of Silenced Echo (ROSE), believing it may give a new start to photon echo in the quantum memory context.

Although three successive pulses are involved in the echo formation, the ROSE scheme strongly differs from what is usually called a "three-pulse photon echo". In the latter process, the atomic coherences resulting from interaction with the first pulse are converted into level population by the second pulse, and restored from population by the third pulse. Instead, all along the present echo process, the captured information is only carried by atomic coherences. The retrieval time of the echo lifts any ambiguity. It equals $t_1+t_{23}+2t_{12}$ in the conventional three-pulse echo, instead of $t_1+2t_{23}$ in the ROSE scheme.     
     
\begin{figure}[h!]
\centering{\includegraphics[width=12cm]{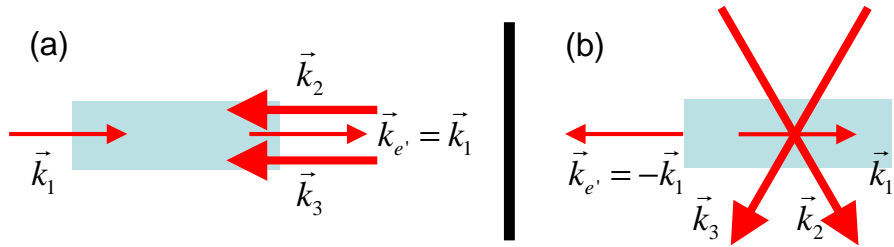}}
\caption{(color online) Beam configuration for signal recovery in forward (a) and backward (b) direction. In (b) the rephasing pulses reach the medium (shaded area) from the sides.}
\label{fig:geometry}
\end{figure}

The CRIB calculations~\cite{sangouard2007} can be extended to ROSE quite directly. As in CRIB, the incoming signal pulse propagates through a spectrally uniform absorbing medium. Moreover, just as in CRIB, the signal is emitted when most atoms sit in the ground state. Hence, when $\vec{k}_3=\vec{k}_2$, the recovery efficiency in forward direction (see Fig.~\ref{fig:geometry}(a))is expected to vary as $(\alpha L)^2\rme^{-\alpha L}$, where $\alpha$ stands for the absorption coefficient, and may reach a maximum value of $54\%$ at $\alpha L=2$, in the absence of coherence relaxation. When the $T_2$  finite duration of the optical coherence is taken into account, the echo intensity can be expressed as $(\alpha L)^2\rme^{-4t_{23}/T_2}$ times the transmitted intensity of the incoming signal, the latter being itself attenuated by factor $\rme^{-\alpha L}$.   

As in CRIB, higher efficiency can be expected when the echo is recovered in backward direction ~\cite{Moiseev2001,kraus2006,sangouard2007,afzelius2009}. This occurs when $(\vec{k}_1,\vec{k}_2)=(\vec{k}_2,\vec{k}_3)=\pi/3$, which leads to $\vec{k}_{e'}=-\vec{k}_1$ (see Fig.~\ref{fig:geometry}(b)). Since they illuminate the storage medium from the sides, the rephasing pulses penetrate along a shorter distance and may undergo less propagation distortion. In the absence of coherence relaxation, the efficiency equals $\left(1-\text{e}^{-\alpha L}\right)^2$ and approaches 100\% when $\alpha L>>1$.   

As in CRIB~\cite{Moiseev2001,kraus2006} or AFC~\cite{afzelius2009,afzelius2010}, two additional $\pi$-pulses may be used to convert the optical coherence into a ground state coherence and back, should a three-level $\Lambda$ system be available. This way the storage time may be increased far beyond the optical coherence lifetime. 

One may wonder whether the absence of primary echo really preserves the information mapping inside the medium. Indeed, when $\left(k_e-k\right)L>\pi$, the echo is killed but radiative emission remains in phase with atomic polarization over distance $\pi/\left(k_e-k\right)$. Therefore some information can creep in the medium over that coherence length. However the mapping is not affected provided $\pi/\left(k_e-k\right)$ is much smaller than the $\alpha^{-1}$ characteristic distance. The radiative migration is radically eliminated in counterpropagating configuration, when $\vec{k}_2=-\vec{k}_1$. Then the coherence length reduces to 1/3 wavelength. When the echo is recovered in backward direction, with $(\vec{k}_1,\vec{k}_2)=\pi/3$, the coherence distance does not exceed one wavelength.    

The ROSE scheme clearly disposes of two insurmountable faults of the conventional 2PE, namely the gain noise and the signal contamination by the FID tail of the rephasing pulses. Spontaneous emission (SE) properties deserve further consideration. During the time interval between the two rephasing pulses, atoms spontaneously decay to the ground state. Those atoms are promoted back to the upper level by the second strong pulse. Therefore, although most atoms sit in the ground state at the moment of the secondary echo, there remains some residual SE noise. In conventional 2PE, when the medium is totally inverted by the rephasing pulse, SE brings a noise of about 1 photon within the time slot and the solid angle of echo emission~\cite{ledingham2010,ruggiero2009}. After a second rephasing pulse, SE from the few atoms left in the upper level brings a small noise contribution to the retrieved signal. Should only 90\% of the atoms be brought back to the ground state, this would still improve the signal-to-noise ratio by a factor of 10 with respect to the 2PE scheme. In addition to those incoherent features, SE is involved in coherent rephasing processes, as discussed in Ref~\cite{ledingham2010}. In the inverted medium, SE can play the role of the first pulse in a 2PE scheme, give rise to an echo after the second rephasing pulse, and spoil the ROSE. However, according to the 2PE phase matching condition, the revived SE noise is radiated close to direction $\vec{k}_3$, at large angular distance from the ROSE.

Finally, the ROSE scheme combines the absence of preparation step with broad-band capability. Unlike the protocols based on a preparation step, such as CRIB, GEM or AFC, the present process does not waste the available optical thickness. All the atoms initially present within the signal bandwidth may participate to the quantum memory. The main spectral width limitation may come from the rephasing pulses. The energy they have to exchange with the medium is indeed proportional to the operation bandwidth. Chirped control pulses offer an advantageous alternative to $\pi$-pulses in terms of bandwidth. In the next section we show they can significantly improve the ROSE rephasing step.

\section{Optimizing the rephasing step}\label{Rephasing pulses}
In the previous discussion, phase reversal is assumed to be achieved by $\pi$-pulses. This may not be the best solution. First, since the spectral range of a $\pi$-pulse is determined by its inverse duration, the pulse energy grows quadratically with the spectral width. Broadband operation demands too much energy. Moreover, when propagating through a broadband absorber, a $\pi$-pulse converts most of its energy into atomic excitation. According to the numerical solution of Maxwell-Bloch equations~\cite{ruggiero2010}, 50\% of a rectangular $\pi$-pulse is absorbed at depth $z=1.78/\alpha$. As a consequence, the pulse undergoes strong distortions. Simultaneous energy loss and area conservation, as prescribed by the McCall and Hahn theorem~\cite{McCall1967}, entail temporal stretching, which corresponds to the narrowing of the excited spectral interval as the pulse proceeds deeper and deeper into the medium.    

Instead of flipping all the Bloch vectors simultaneously, one may rotate them sequentially with the help of chirped pulses, whose frequency is scanned at rate $r$. Under the conditions of Adiabatic Rapid Passage (ARP), the Bloch vector adiabatically follows the driving vector on the Bloch sphere and $\pi$-radian flipping can be achieved efficiently over the scanning range of the driving field. Unfortunately, because atoms with different frequencies are excited successively, such an ARP process cannot refocus the Bloch vectors and give rise to an echo. More precisely, a $\tau$-long incoming signal cannot be recovered in the shape of a $\tau$-long echo, except if the chirped pulse behaves as a brief excitation, interacting with all the atoms quite simultaneously. Therefore, since the scanning time of the initially excited spectral interval is given by $2\pi/(r\tau)$, the Bloch vectors can be indeed refocused provided $2\pi/(r\tau)<<\tau$, a situation of little use where ARP does not really differ from $\pi$-pulse excitation. More interesting is the excitation by a \textit{pair} of chirped pulses. Then the atomic coherences can be phased together~\cite{lauro2011}, whatever the size of $\phi=2\pi/(r\tau^2)$. This precisely corresponds to the scheme we are considering. 

The simplest ARP process is obtained by chirping the frequency of a field with fixed amplitude at constant rate. However the field has to be switched on and off at large detunings from the atoms. Finite duration excitation gives rise to oscillatory features~\cite{vitanov1996}. To flip the Bloch vector over a well defined finite spectral interval, a Complex Hyperbolic Secant (CHS) pulse is more appropriate~\cite{roos2004}. High transfer efficiency has been reached over the targeted spectral band, while the atoms are left at rest outside~\cite{deSeze2005,rippe2005}.  

Frequency chirped pulses seem to be able to keep their properties during propagation through an absorbing material~\cite{zafarullah2007}. Specifically, they lose a much smaller fraction of energy than a $\pi$-pulse. To illustrate this property, let us consider a CHS pulse. The Rabi frequency and instantaneous frequency temporal variations are respectively defined by:
\begin{align}
\Omega(t)&=\Omega_0\text{sech}\left(\beta(t-t_0)\right)\label{eq:CHS1}\\
\omega(t)&=\omega_0+\mu\beta\tanh\left(\beta(t-t_0)\right)\label{eq:CHS2}
\end{align} 
According to Eq. \ref{eq:CHS2}, the field is chirped over a $2\mu\beta$-wide interval centered at $\omega_0$. The chirp rate $\text{d}\omega(t)/\text{d}t$ reaches its maximum value $r_0=\mu\beta^2$ at $t=t_0$. When the atoms are promoted to the upper level, the energy conveyed to the medium reads as:
\begin{equation}
W_{at}=2\mu\beta ALn\hbar\omega_0
\end{equation}
where $L$, $A$, and $n$ respectively represent the medium length and cross-sectional area, and the spectro-spatial atomic density. The incoming energy reads as:
\begin{equation}
W_{in}=\frac{1}{2}Ac\epsilon_0\int\text{dt}|\Omega(t)\hbar/d|^2=Ac\epsilon_0|\Omega_0\hbar/d|^2/\beta  
\end{equation}     
where $d$ represents the transition dipole moment. The fraction of incoming energy that is stored in the medium can be expressed in terms of the absorption coefficient $\alpha=\pi\omega_0d^2n/(c\hbar\epsilon_0)$ as:
\begin{equation}
\frac{W_{at}}{W_{in}}=\frac{2}{\pi}\frac{r_0}{\Omega_0^2}\alpha L  
\end{equation}
Since adiabatic passage precisely requires that $r_0/\Omega_0^2<<1$~\cite{deSeze2005}, the storage of a small fraction of energy in the medium can be made consistent with the adiabatic condition, whatever the optical depth $\alpha L$.  

%- un seul PAR ne produit un bon écho qu'à condition que delta²<<r, situation où on rejoint les propriétés d'une impulsion \pi et qui n'est donc pas favorable.
%- Noter que la situation intéressante correspond à µ>>1.
\section{Experimental}\label{experimental}  
Experimentally we pursue a twofold objective. First we need to confirm the generation of the revived echo, with the expected efficiency, when the primary echo is silenced by spatial phase mismatching. We concentrate on the counterpropagating configuration (see Fig.~\ref{fig:geometry}(a)) where mismatching is maximum. The two rephasing pulses follow the same path, in opposite direction to the signal field. Second we want to verify that the second rephasing pulse actually brings the atoms back to the ground state. 

A first experiment is conducted in a $0.5\%$ at. Tm$^{3+}$:YAG crystal operating at 793nm. The sample is cooled down to 2.8K in a variable temperature liquid helium cryostat. The counterpropagating light beams are split from a home-made, extended cavity, continuous wave, ultra-stable semi-conductor laser~\cite{crozatier2004}, featuring a stability better than 1 kHz over 10 milliseconds. To manage the rephasing and signal paths separately, we make a small angle of $\approx2^o$ between them. This way we not only extract the echo easily but also further reject reflections from the cryostat windows. Those precautions will prove essential in shot-noise limited measurements.   

\begin{figure}[h!]
\centering{\includegraphics[width=10cm]{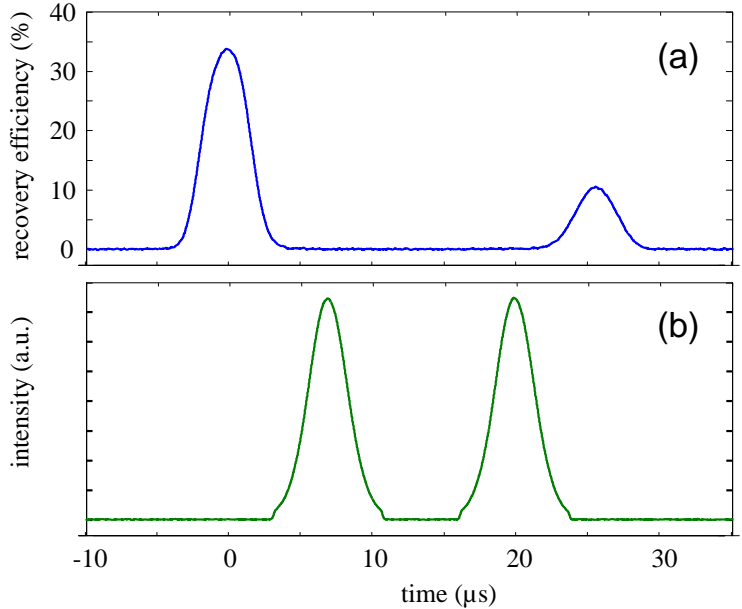}}
\caption{(color online) Revival Of Silenced Echo in Tm$^{3+}$:YAG. Opacity is adjusted to $\alpha L=1.05$. The incoming pulse and the echo are displayed in (a). They counterpropagate with the CHS rephasing pulses, detected in the opposite direction, and represented in (b). In (a), efficiency refers to the intensity of the incoming signal, measured at the input side of the crystal.}
\label{fig:TmYAG}
\end{figure}

The signal beam waist is adjusted to 30$\mu$m, with a rephasing beam 1.6 times bigger. All the fields are polarized along the $[001]$ crystallographic axis~\cite{sun2000}. In order to avoid propagation effects, we slightly detune the laser from the absorption line center to $\alpha L=1.05$. 

The rather high Tm$^{3+}$ concentration reduces $T_2$ to less than 50$\mu$s, making the ROSE wither fast. This impacts on the parameter accessible range. To simultaneously satisfy $T_2\beta>>1$ and  $\mu\beta^2<<\Omega_0^2$ (adiabatic passage condition), one is led to the minimum $\mu=2$ acceptable value, with $\beta/(2\pi)=120$kHz. Hence population is inversed over a $\mu\beta/\pi=480$kHz-wide spectral interval.  

With a rephasing pulse separation of $t_{23}=13\mu$s, the storage time equals $26\mu$s. The $3.5\mu$s-long (full width at half maximum) incoming signal and the recovered ROSE are displayed in Fig.~\ref{fig:TmYAG}(a). Spatial-mode filtering through a single mode fiber at the crystal output efficiently rejects stray light from the the CHS rephasing pulses. Those pulses are monitored with a control detector, as shown in Fig.~\ref{fig:TmYAG}(b). As can be seen in this figure, we truncate the CHS waveforms to 6$\beta^{-1}$. The recovery efficiency of $\approx10\%$ is consistent with the $(\alpha L)^2\rme^{-\alpha L-4t_{23}/T_2}$ predicted value, setting $T_2=42\mu$s. The latter $T_2$ value agrees with independent measurements. 

The refocusing parameter $\phi=2\pi/(r\tau^2)$ defined in Section~\ref{Rephasing pulses} reads as $\phi=2\pi/(\mu\beta^2\tau^2)$. To accomodate the incoming pulse bandwidth, CHS parameters must satisfy $\mu\beta\tau>\pi$, which results in $\phi<2\mu/\pi$. With $\mu=2$, $\phi$ remains smaller than unity, thus allowing for some refocusing of the Bloch vectors after the first CHS pulse. In order to reach larger $\phi$-values, we undertake a second experiment in a $0.005\%$ at. Er$^{3+}$:YSO, $7.5$mm-thick, crystal at 1.5$\mu$m. This material offers attractively slow relaxation rates that let time for the ROSE to blooming. The upper level decays in $10$ms and we have measured an optical coherence lifetime $T_2$ as large as $230\mu$s. To reach such a high $T_2$ value, one cools down the crystal to $2$K and one lifts the Kramers degeneracy with a magnetic field of $\approx2.2$T. The light beams are directed along the $b$ axis, perpendicular to the magnetic field. The crystal is birefringent and $T_2$ is optimized when the neutral lines $D_1$ and $D_2$ are oriented at $45^o$ from the magnetic field ~\cite{bottger2002}. 

The light source is an erbium-doped fiber, distributed feedback, commercial laser (KOHERAS). This source features good stability, with a coherence time larger than 1ms and a jitter of a few kHz over 100$\mu$s. The rephasing beam is 2.5 times bigger than the signal beam, whose beam waist is measured to be $45\mu$m at the crystal. All the fields are polarized along $D_1$. The opacity $\alpha L$ reaches $\approx2$ at the line center. The echo is extracted by a beam splitter and collected on a photodiode.    
\begin{figure}[h]
\centering{\includegraphics[width=10cm]{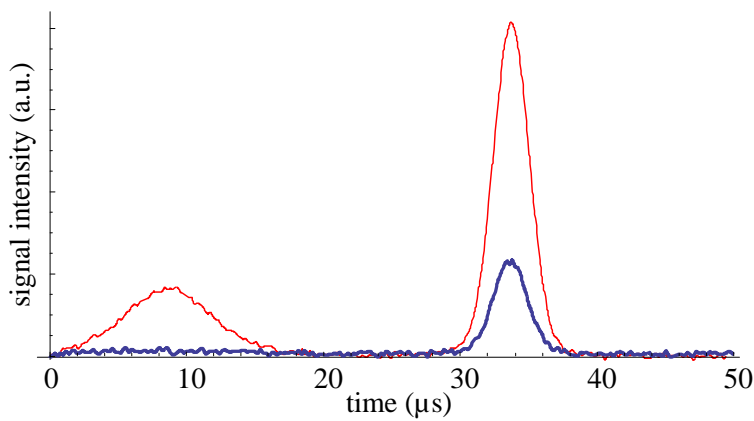}}
\caption{(color online) Population inversion by a CHS pulse. A probe pulse is detected before (thick blue line) and after (red thin line) excitation by a $400$kHz-wide CHS pulse. Some stray light from the cryostat windows reveals the CHS profile and time position.}
\label{fig:ampli_check}
\end{figure}

First we check the inversion efficiency of a CHS pulse with parameters $\beta=125\:10^3\text{s}^{-1}$ and $\mu=10$, corresponding to excitation over a $400$kHz-wide interval. As before, the CHS duration is limited to 6$\beta^{-1}$. Slightly detuning the laser from the line center, we measure an opacity $\alpha L=0.71$. A $3\mu$s-long (full width at half maximum) gaussian-shaped signal pulse is shone through the crystal. We successively detect the transmitted profile before and after inversion by a counterpropagating CHS pulse (Fig.~\ref{fig:ampli_check}). The pulse sequence is repeated at $10$Hz-rate. We observe an amplification factor of $3.55$. Through a totally inverted medium, with an upper-level population $n_b=1$, this factor should reach $\text{e}^{2\alpha L}=4.13$. The measured factor corresponds to $n_b=\text{Ln}(3.55)/(2\alpha L)=0.89$.
\begin{figure}[h]
\centering{\includegraphics[width=10cm]{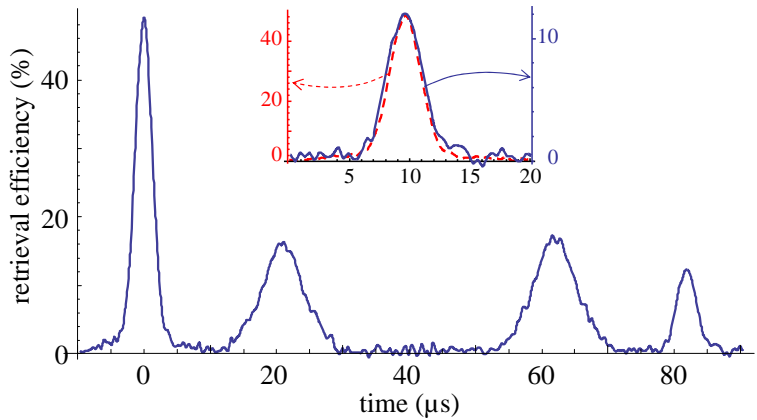}}
\caption{(color online) ERevival of silenced echo in Er$^{3+}$:YSO. After incoming of a 3$\mu$s-long weak pulse, two CHS pulses successively illuminate the crystal from the opposite direction. They cover a 0.4MHz wide spectral interval. Their time separation is adjusted to 41$\mu$s. Their time evolution is revealed by reflexions from the cryostat windows. The echo is radiated 82$\mu$s after the first pulse. The incoming signal (dashed line) and echo (solid line) temporal profiles are compared in the inset.}
\label{fig:secondary_echo}
\end{figure}   

With the same CHS parameter values, the same input signal duration and the same opacity, we check the ROSE process. In the experiment presented in Fig.~\ref{fig:secondary_echo}, the rephasing pulse separation is adjusted to $41\mu$s, which leads to a storage time of $82\mu$s. The recovery efficiency, as measured with respect to the incoming pulse at the crystal input, reaches $12\%$, in excellent agreement with the $(\alpha L)^2\rme^{-\alpha L-4t_{23}/T_2}$ expected value (see section~\ref{protocol}). As shown in Fig.~\ref{fig:secondary_echo} inset, the temporal profile of the echo nearly coincides with that of the incoming pulse, showing that the CHS spectral range is correctly adjusted to the signal width. The refocusing parameter $2\pi/(\mu\beta^2\tau^2)$ reaches $\approx4.5$, attesting to the absence of Bloch vector refocusing after the first CHS pulse.  
\begin{figure}[h]
\centering{\includegraphics[width=10cm]{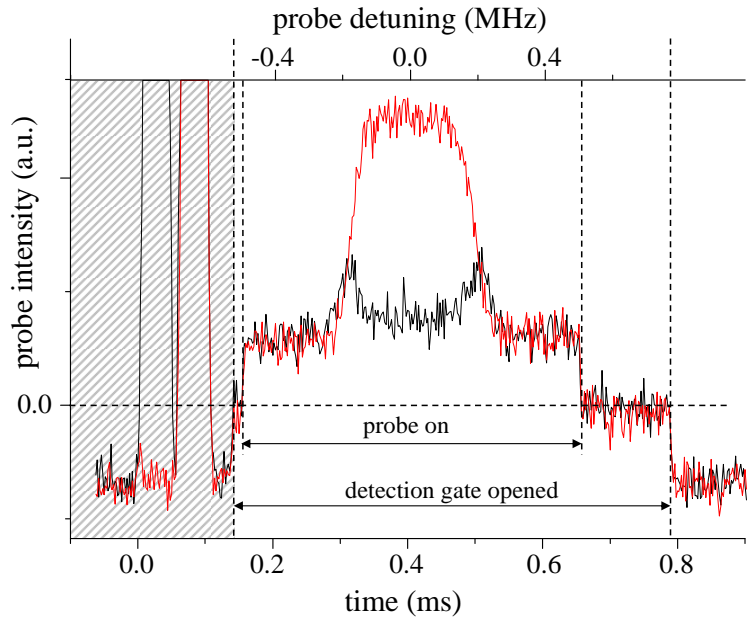}}
\caption{(color online) The transmission spectrum is probed after illumination by one (red line) or two (black line) CHS pulses. The probe pulse is scanned over 1MHz in 0.5ms. The upper level population reaches 0.7 after one CHS and drops to less than 0.1 after the second CHS. The opacity is adjusted to $\alpha L=1.0$. }
\label{fig:transmission_spectrum}
\end{figure}   

To check the CHS ability to bring ions back to the ground state, we probe the crystal transmission spectrum after illumination by one or two CHS pulses. In order to optimize the beam overlap, we slightly modify the setup. All the light fields now copropagate, being carried by a single beam. The detector is protected from the intense CHS pulses by an acousto-optic modulator operating as an optical switch. The transmitted light is detected through a pinhole that selects the uniformly excited central region of the crystal. The CHS pulses are defined by the same $\beta$ and $\mu$ values as above. The Rabi frequency is adjusted to $\approx3.5\:10^6$s$^{-1}$. The probe field is scanned over a 1MHz-wide interval in 0.5ms. The recorded spectra are displayed in Fig.~\ref{fig:transmission_spectrum}. The inverted interval edges are not vertical. Instead they exhibit a finite width of $\approx50$kHz, consistent with the computed value obtained with $\mu$=10 and a finite pulse duration of $6\beta^{-1}$. The laser being tuned to $\alpha L=1.0$, the upper level population reaches 0.7 after one CHS and drops to less than 0.1 after the second CHS. As discussed in Section~\ref{protocol}, this should be enough to reduce the spontaneous emission noise by 10 with respect to the conventional 2PE scheme.        
   
\section{Conclusion}
We have described an original Revival Of Silenced Echo (ROSE) scheme where an echo signal is radiated from ground state atoms with optimum efficiency. This process remedies the conventional photon echo drawbacks~\cite{longdell2010,ruggiero2009,sangouard2010} in the prospect of quantum storage. In the small optical thickness regime, optimum recovery efficiency has been reached experimentally. Complex hyperbolic secant pulses, more robust to propagation than $\pi$-pulses, have proved efficient in rephasing the atomic coherences. With respect to previously proposed schemes such as CRIB and AFC, ROSE avoids complex and optical-depth consuming preparation steps, while preserving the photon echo broadband and multimode capability.  

We are pleased to acknowledge stimulating and fruitful discussions with J. J. Longdell.

This work is supported by the European Commission through the FP7 QuRep project, by the national grant ANR-09-BLAN-0333-03 and by the Direction G\'en\'erale de l\'\ Armement.

\end{document}